\documentclass{caps}
\usepackage{peter}
\usepackage{epsf}

\def\apj{\textit{Astrophysical Journal}}

\def\aap{\textit{Astronomy \& Astrophysics}}

\def\mnras{\textit{Monthly Notices of the Royal Astronomical Society}}
\def \lta {\mathrel{\vcenter
     {\hbox{$<$}\nointerlineskip\hbox{$\sim$}}}} 

\def\gcm3{g~cm$^{-3}$}
\def\g-s{g~s$^{-1}$}
\def\cm3s{cm$^3$~s$^{-1}$}

\def\erg-s{erg~s$^{-1}$}

\def\beq{\begin{equation}}
\def\eeq{\end{equation}}

\begin{document}

\author{J. Craig Wheeler\\Department of Astronomy\\
University of Texas at Austin
}

\chapter{Conference Summary: Three-Dimensional Explosions
}


\section{Introduction}

This conference was packed with interesting and relevant
developments regarding the three-dimensional nature of
both thermonuclear and core-collapse supernovae.  Before
summarizing those presentations, I would like to summarize
some of the developments regarding rotation and magnetic
fields that were on my mind during the conference.

\section{Dynamo Theory and Saturation Fields}

There has been a major breakthrough in the conceptual
understanding of astrophysical dynamos in the last few years.
In traditional mean field dynamo theory, the turbulent velocity
field that drives the ``alpha" portion of the $\alpha - \Omega$
dynamo was specified and held fixed.  A weakness of the
original theory was that the turbulent velocity field cannot be
constant. The buildup of small scale magnetic field tends to inhibit
turbulence, cutting off the dynamo process for both small and
large scale fields. Since the small scale field tended to grow faster
than the large scale field, it appeared that the growth of the large
scale field would be suppressed (Kulsrud \& Anderson 1992;
Gruzinov \& Diamond 1994).  In these theories, the magnetic field energy
cascades to smaller length scales where it is ultimately dissipated
at the resistive scale.  Large scale fields tend to build up slowly,
if at all.

The solution to this problem has been the recognition (Blackman \& Field
2000; Vishniac \& Cho 2001; Field \& Blackman 2002;
Blackman \& Brandenburg, 2002; Blackman \& Field 2002;
Kleeorin et al. 2002) that the magnetic helicity, {\bf H = A$\cdot$B}
is conserved in ideal MHD and that this conservation had not been
treated explicitly in mean field dynamo theory.  Incorporation of
this principle leads to an
``inverse cascade" of helical field energy to large scales that is
simultaneous with the cascade of helical field energy from the driving
scale to the dissipation scale.  Basically, the large scale helical field
and inverse cascade must exist with opposite magnetic helicity to that
of the field cascading to small scale.  The result (Blackman \&
Brandenburg 2002) is the rapid growth of large scale field in a kinematic
phase (prior to significant back-reaction) to a strength where the
field on both large and small scales is nearly in equipartition with
the turbulent energy density.  At that point, the back reaction sets
in and there tends to be a slower growth to saturation at field
strengths that can actually somewhat exceed the turbulent energy density.
It may be that the early, fast, kinematic growth is the only phase that
is important for astrophysical dynamos, especially in situations that
have open boundaries so that field can escape (Brandenburg, Blackman \&
Sarson 2003; Blackman \& Tan 2003) and that are very dynamic.
The collapse ambience is clearly one of those situations.

        Another possibly important insight is that the rapid kinematic
phase can lead to magnetic helicity currents (Vishniac \& Cho 2001).
It is possible that these magnetic helicity currents can
transport power out of the system in twisting, propagating
magnetic fields. This is clearly reminiscent of jets or winds,
but the physics is rather different than any that has been
previously explored  in driving jets or winds.  This physics
needs to be explored in the context of supernovae and
gamma-ray bursts.

This new work on dynamo theory has not changed one basic aspect
and that is the level of the saturation fields.  It remains true
that the saturation fields will be of order v$_{\mathrm{a}}\sim\lambda\Omega$
or B$^2\sim 4\pi \rho {\lambda^2} \Omega^2$ where the characteristic wavelength, 
$\lambda \lta$ r, for quasi-spherical situations.  For a proto-neutron star 
this yields a field of order 10$^{15}$ to 10$^{16}$ G.
For collapse to form a black hole, the velocities will be
Keplerian and the associated, dynamo-driven, predominantly
toroidal field will have a strength of order B$\sim$10$^{16}$G
$\rho_{10}^{1/2}$ assuming motion, including the Alfv\'en speed,
near the speed of light near the Schwarzschild radius and a
characteristic density of order 10$^{10}$ g cm$^{-3}$
(MacFadyen \& Woosley 1999).  Fields this large could affect both
the dynamics  and the microphysics in the black hole-formation
problem.  Because of the nearly Keplerian motion in the black
hole case, the fields generated will be much closer to pressure
equipartition than in the neutron star case, and hence, perhaps,
even more likely to have a direct dynamical effect. The
associated MHD power in the black hole case would be roughly
$10^{52}-10^{53}$ erg s$^{-1}$.

\section{Possible Effects of Large Magnetic Fields}

\subsection*{A.  Equation of State}

Fields of order 10$^{15}$  to 10$^{16}$  G are far above the QED
limit, B$_{\mathrm QED} = 4\times10^{13}$ G, so quantum effects
may become important.  The calculations of
Akiyama et al. (2003) predict regions $\sim 10^6$ to $10{^7}$  cm
where the electron Fermi energy is less than the first Landau
level after about 100 ms (see the contribution in these proceedings
by Akiyama et al.).
In such conditions, electron motions will be quantized, with the electron
component  of the pressure being strongly anisotropic.  This
pressure anisotropy is likely to be balanced by the {\bf j $\times$ B}
force of induced magnetization (Blandford \& Hernquist 1982), but in
the absence of such isotropy, pressure anisotropy of order
10$^{-4}$  and hence velocity anisotropy of order $10^{-2}$  might
be induced.  The electron pressure will be reduced compared to
calculations that ignore quantization, but it is not clear that
will make a significant difference to the dynamics.


For B $>$ B$_{\mathrm QED}$, the electrons can only
flow along the field lines, that is \textbf{j}  $||$ \textbf{B}.
On the other hand, classic MHD includes currents only
implicitly and assumes that the current is always normal to the
field, \textbf{j} $\perp$ \textbf{B}.  The result is a manifest
contradiction,
as pointed out to me by Dave Meier.  The resolution to this might
be non-local currents, ion currents (which would require flows
of only 10$^{-6}$ cm s$^{-1}$), or most interestingly, but
unlikely, a field that saturates at the QED limit.  These issues
are worth more thought.

\subsection*{B. Neutrino Transport}

Fields of order 10$^{15}$ to 10$^{16}$ G that will characterize
both neutron star and black hole formation may affect neutrino
transport. With a large magnetic field, direct $\nu-\gamma$ interaction is
possible mediated by W and Z bosons.  This would allow
neutrino Cerenkov radiation, $\nu \rightarrow\nu +
\gamma$, and would enhance plasmon decay, $\gamma \rightarrow \nu + \nu$
(Konar 1997).

In addition, processes like $\nu \rightarrow\nu + \mathrm e^+ +\mathrm e^-$
would no longer be kinematically forbidden.  In that case, closed
magnetic flux loops can trap pairs.  The energy in pairs would
grow exponentially  to the point where annhilation cooling would
balance pair creation.  Thompson \& Duncan (1993) estimated that
an energy as much as E$_{\mathrm{pair}}\sim 10^{50}$ erg could be
trapped in this way.  This is not enough energy to cause a robust
explosion, but it is enough energy to drive the dynamics of core
collapse in a substantially different way, perhaps by inducing
anisotropic flow if the flux loops are themselves distributed
anisotropically.

With substantial magnetic fields, the cross section for inverse
beta decay, $\nu_{\mathrm e} +\mathrm n\rightarrow\mathrm p +
\mathrm e^-$, would become dependent  on neutrino momentum,
especially for asymmetric field distributions, which would be
the norm (Lai \& Qian 1998; Bhattacharya \& Pal 2003; Ando 2003).

All these processes and more should be considered quantitatively
in core collapse to form neutron stars and black holes.

\section{Core Collapse MHD and Jet Formation}

\subsection*{A.  Magnetic Helicity Currents}

It is not at all proven that the large magnetic fields expected
in core collapse generate  jets, but there are a number of clues
pointing in that direction.  For the more traditional situation
in which collapse leads to the formation of a neutron star, the
premise is that there is a rapid formation of a strong magnetic
field with B$\sim$ 100 B$_\mathrm{QED}<< (4 \pi \mathrm
P)^{1/2}$, that is much above the QED limit, but less than
equipartition with the ambient pressure.  This field is expected
to be primarily toroidal (simulations give $\sim$80$\%$; Hawley
Gammie \& Balbus 1996), but turbulent, with a maximum around the
proto-neutron star surface, a location well within the standing
shock.  The expected MHD power, $\sim 10^{52}$  erg s$^{-1}$,
would be delivered in some form beneath that shock and could help
to reinvigorate  it, or to provide entirely unique, jet-like
dynamics in which the shock no longer played a key role.
In this highly magnetized environment, there will be hoop
stresses, gradients in magnetic  pressure and perhaps in the
electron pressure.  These anisotropic components will be weak
compared to the total pressure, but they will be non-radial and
anisotropic.

As an example of the possibly relevant physics,
Vishniac \& Cho (2001) argue that along with conservation of
magnetic helicity, {\bf H = A $\cdot$ B}, and the inverse
cascade of magnetic field energy to large scales, one will get a
current of magnetic helicity that can be crudely represented by
\begin{equation}
 J_{\mathrm H}\sim \mathrm B^2\lambda \mathrm v,
\end{equation}
where the characteristic length, $\lambda$, might be comparable to a 
pressure scale height, $\ell_{\mathrm P}$ = (d ln P/dr)$^{-1}$, and 
v$\sim \mathrm v_{\mathrm a}\sim \ell_{\mathrm P}\Omega$.  The energy 
flux associated with this magnetic helicity current is J$_{\mathrm H}/ 
\lambda\sim\mathrm B^2\mathrm v_{\mathrm a}$, and so with $\mathrm 
B^2 \sim \rho \ell_{\mathrm P}^2 \Omega^2$ the associated power is:
\begin{equation}
\mathrm L =\mathrm r^2\mathrm B^2\mathrm v_{\mathrm a}\sim
\mathrm B^2\mathrm r^2\ell_{\mathrm P}\Omega\sim\rho\mathrm r^5\Omega^3
\left(\frac{\ell_{\mathrm P}}{r}\right)^3.
\end{equation}
Note that the next-to-last expression on the RHS is essentially
just the characteristic Blandford-Payne luminosity
(Blandford \& Payne 1982); however, in this case the field is not
externally given, but
provided by the dynamo process so that the final expression on
the RHS is given entirely in terms of local, internal
quantities.  The implication is that this amount of power is
available in an axial, helical field without twisting an
external field. Again, while this analysis has superficial
resemblance to other jet mechanisms, it involves rather
different physics and is self-contained.
Whether this truly provides a jet remains to be seen.  A first
example of driving a polar flow with the MRI is given by Hawley \&
Balbus (2002).

Note that this process of creating a large scale field with an
MRI-driven dynamo with its promise of naturally driving axial,
helical flows does not require an equipartition field.  As pointed
out by Wheeler et al. (2002), the field does not have to have
equipartition strength and hence to be directly dynamically
important in order to be critical to the process of core
collapse.  The field only has to be significantly strong to
catalyze the conversion of the free energy of differential
rotation of the neutron star into jet energy.  As long as this
catalytic function is operative, the rotational energy should be
pumped into axial flow energy until there is no more
differential rotation.  For the case of stellar collapse, this
would seem to imply that, given enough rotational energy in the
neutron star,  this machine will work until there is a
successful explosion.  Even if the core collapses directly into
a black hole, or does so after some fall-back delay, the basic
physics outlined here, including magnetic helicity currents and
their associated power should also pertain to black hole formation.

\subsection*{B.  Poleward Slip Instability}

        Another interesting bit of physics that may pertain to core
collapse is the poleward slip instability.  This is analogous
to wrapping a rubber band around the equator of a ball and then
sliding it upward.  For the case of a magnetized plasma, a
toroidal field is absolutely unstable to this effect in the
absence of rotation (Spruit \& Ballegooijen 1982).  The case with
differential rotation has been considered by, among others,
Chanmugam (1979).  In that case the axisymmetric (m = 0) mode
still appears to be unstable, but this case is a bit tricky
because the absence of a sufficient condition for stability as
derived by Chanmugam does not necessarily imply a necessary and
sufficient condition for instability.  The interesting behavior,
in any case, is not merely the linear instability, but the
non-linear dynamics.  This does not seem to have been explored
at all in the literature.

As a crude way of examining this,
let us assume that the pressure gradient balances gravity to first order
and look at the acceleration resulting from the hoop stress and centrifugal
potential, assuming conservation of angular momentum of the matter
associated with a flux tube.  The result is
\begin{equation}
\mathrm a \sim \frac{\mathrm{v_a}^2}{\mathrm r} -
\frac{\mathrm R^4\Omega_{\mathrm eq}^2}{\mathrm r^3},
\end{equation}
where r is the cylindrical radius, R the value on the equator,
and $\Omega_{\mathrm eq}$ the value of the angular velocity on the
equator.  For the case of interest, the saturation field condition 
is that v$_{\mathrm a}\sim$R$\Omega_{\mathrm eq}$, so that these 
terms nearly cancel on the equator.  This is a caution, at least, that 
care must be taken to take all the forces into account self-consistently.
The issue of what happens as the field starts to slip toward the
pole seems to depend on the behavior of the Alfv\'en velocity, and
hence the magnetic field and entrained density, as the flux tube
moves. 

Note that the poleward slip instability, whatever its ultimate
non-linear behavior, should not depend on whether the field is
continuously connected around the body (literally like a rubber
band) or whether it is turbulent and discontinuous.  This is
because, for instance, the hoop stress is a local property of a
field with a mean radius of curvature.  Williams (2003 and this
conference) has argued
that  even a tangled field with $< $B $> \sim$ 0, but $< \mathrm
B^2 >^{1/2} \neq$ 0, will act like a viscoelastic fluid (see
also Ogilvie 2001) and, in particular, exert a hoop stress.

The conjecture is that the ultimate non-linear behavior is for
the field to accumulate near the pole where it reaches
approximate equipartition, B $\sim (8 \pi \mathrm P)^{1/2}$, and
hence becomes dynamically significant.  Again, this suggests
activity at the pole that is reminiscent  of a jet.  Yet again,
this remains to be seen.

One interesting aspect of the poleward slip instability is that
it would seem to pertain directly to neutron stars that have a
strong density gradient at the surface, essentially a hard
surface, but it should \textit{not} work for black holes, where there is
no surface to support the poleward slip.  Whether or not this makes any
difference in the jet formation in neutron stars versus black
holes is an interesting question.

\section{Summary of Contributions}

\subsection{Asymmetry Rules}

The conference began with excellent summaries of the new and
growing sample of supernova spectropolarimetry by Lifan
Wang and Alex Filippenko.  It is this data that has driven
the new conviction that core collapse supernovae are essentially
universially asymmetric and that the asymmetry is driven by
the engine of core collapse itself.  With this new conviction,
disparate data on otherwise isolated events like the Crab nebula
with its pulsar and jet, Cas A, and SN~1987A, begin to make sense
in a large picture of fundamentally asymmetric supernovae.
Roger Chevalier discussed our evolving knowledge of supernova
remnants and pulsar wind nebulae. Rob Fesen described observations
on the morphology of supernovae remnants, especially Cas A.
Bob Kirshner summarized the imaging spectroscopy on SN~1987A.
Doug Swartz described the new data on supernova remnants
available from the Chandra Observatory.  Vikram Dwarkadas showed
that his multidimensional simulations of supernova ejecta colliding
with previously expelled wind material are rife with
Rayleigh-Taylor instabilities and look remarkably like the observations.

One of the lessons that comes through from this
work is that neither Type Ia nor the zoo of core collapse
supernovae are spherically symmetric.  Peter H\"oflich and
the fully-represented Oklahoma mafia -- David Branch, R. C. Thomas,
Dan Kasen, and Eric Lenz, with Eddie Baron kibbutzing from the
audience -- outlined the various ways in which polarization
could be induced in supernova spectra.  Among these are:
an intrinsically asymmetric shape, blocking of part of the
photosphere by some off-center distribution of matter, and
an off-center energy source.  All of these may contribute in
various supernovae or even for a single supernova, depending
on circumstances.
 
\subsection{Type Ia}
 
As mentioned in my introduction, one of the goals of the study
of Type Ia supernova research for decades has been to obtain
direct observational evidence that Type Ia arise in binary
systems, as widely accepted on circumstantial grounds. This
conference may have revealed some of the first evidence
in this direction.  Lifan Wang, Peter H\"oflich, and Dan Kasen
discussed the observations and interpretations of polarization data
from Type Ia supernovae, particularly the ``normal" event
SN~2001el that shows remarkable departure from symmetry in the form
of a highly polarized high-velocity component to the Ca II IR triplet.  
After the conference, Gerardy et al. (2003) submitted a paper arguing
that a similar high-velocity Ca feature in SN~2003du might arise
in a hydrogen-rich circumstellar medium. The data have not yet revealed 
definate proof, but tantalizing suggestions that the asymmetry may be 
connected to a disk or binary companion, the existence of which 
would be proof that a binary system was needed. 

Mario Hamuy added a dramatic new development in this area with
his discussion of SN~2002ic, an event that shows familiar Type Ia
features, but also strong hydrogen emission lines similar to those
from Type IIn.  A substantial amount of hydrogen, of order a
solar mass, must be involved. After the conference, Wang et al. (2003)
submitted a paper based on VLT spectropolarimetry observations
that showed that the hydrogen envelope is substantially polarized
and probably arrayed in a large, dense, clumpy disk-like way.
SN~2002ic is very similar both near maximum light and 200 days
later to SN~1997cy and SN~1999E, both classified as Type  IIn.
This raises the issue of whether or not at least some of these
events previously classified as SN IIn are hydrogen-surrounded
Type Ia.  These events are rather
rare, so it cannot be true that all Type Ia erupt in this
configuration, but it is also clear that Hamuy has provided us
with a stimulating new avenue of exploration of the nature
of Type Ia and their binary configuration.

As a complement to this, Don Winget described the work that he
and his group are doing with asteroseismology to probe the inner
composition of white dwarfs.  Sumner Starrfield and S.C. Yoon gave very
thought-provoking summaries of their work that gives new insights
into the possible configurations of white dwarf accretion and
growth that could lead to Type Ia explosions.
Jim Truran and Andy Howell both provided insights into how the
diversity of Type Ia supernovae may arise.

There has been amazing progress on understanding and simulating the
combustion physics associated with Type Ia thermonuclear explosions,
as summarized by Alexei Khokhlov, Elaine Oran, Vadim Gamezo, Eli Livne,
and Peter H\"oflich.  In particular Gamezo illustrated the state of the
art with a three-dimensional simulation of a detonation
that starts deep in the fingers of unburned carbon and oxygen that
survive at the end of the phase of subsonic, turbulent, deflagration.
Fundamental understanding of the deflagration/detonation transition in
this ``unconfined" problem may be just around the corner.

We also had summaries of the dramatic application of Type Ia supernovae
to cosmology and the prospects for probing the ``dark energy" from
Brian Schmidt and Saul Perlmutter.  The astounding discovery of the
acceleration of the Universe did not depend on any deep understanding
of the physics of the explosion, nor on the evidence for asymmetry
being revealed by spectropolarimetry.  As we try to measure the
effective acceleration as a function of space and time, effectively
the equation of state of the dark energy, systematic effects must
be mastered at an unprecedented level of precision.  This will
require a greater physical understanding and an understanding of the
origin of the asymmetries that may give a dependence of the luminosity
on the angle of observation.  If all Type Ia are basically alike,
then such angle-dependent effects will average out in a large sample,
but if, for instance, the cause of the asymmetry varies with
redshift because the underlying cause of the asymmetry does,
then great care will be required to make the appropriate analysis
of the high-redshift observations.

\subsection{Core Collapse}

To emphasize, the lesson that emerges strongly from
recent studies of the polarization of supernovae and related
issues is that core collapse supernovae are always asymmetric,
and frequently, but not universally, bi-polar.  I must emphasize
that this is a hard won conclusion, with heroic observational work
by Lifan Wang and by Doug Leonard, Alex Filippenko and their
colleagues.  

In terms of giving credit, Stirling Colgate revealed
the true father of modern supernova research:
Scratchy Serapkin.  Anatoly Serapkin was the head of the Soviet
delegation to the Geneva talks aimed at the Limited Test Ban Treaty
to abandon space, atmospheric, and underwater nuclear talks in  1963.
Colgate was one of the
representatives on the U.S. side with the self-appointed goal of
convincing both sides that we needed to understand the astrophysical
``background" to avoid confusing a natural event with a bomb test.
The yet-to-be famous Vela satellites played a role in these discussions.
Colgate said supernovae might be confused with a test.  Scratchy, not
a scientist himself, fixed him with a steely glare and inquired,
``Who knows how supernovae work?"  Colgate realized what thin ground
he, and the U.S. delegation, were on, returned to Livermore and
made the case to Edward Teller that understanding supernovae must
become a primary goal of the lab.  The rest is history.

At this conference, new perspectives on the mechanisms
of core collapse, neutrino transport, rotation, and  magnetic fields
were given by Stirling Colgate, Adam Burrows, Thiery Foglizzo, Dave Meier,
Dong Lai, and Peter Williams.
The impact of asymmetries on the dynamics and on the question
of neutron star versus black hole formation were also discussed.
Issues of nucleosynthesis were discussed by John Cowan, Keichi Maeda,
and Raph Hix.

Mario Hamuy also spoke about the large range in apparent kinetic
energy of the explosions of Type II supernovae.  He raised the
question of whether or not the distribution of 
energy is a continuum,
or is more complex, implying,
perhaps, different physical processes. An example would be neutron
star versus black hole formation.  These questions must also be
posed for Type Ic supernovae.  We need to determine if the events
labeled ``hypernovae" by Maeda and his collaborators are truely
special, or part of a continuum.  Given the evidence for strong
asymmetries, there will be line-of-sight effects early on.  Asymmetric
flows are also apt to alter the systematics of gamma-ray deposition
in later phases, and hence the luminosity and slope of the radioactive tail.
Alejandro Clocchiatti reported some old, but still quite relevant,
data on Type Ic events that may help to resolve such issues.

Norbert Langer argued that the variations in single star and
binary evolution make it unlikely that all massive stars will
undergo the same rotational history.  Thus, it is improbable to
have all iron cores evolve with the same rapid rotation prior to
collapse as may be required for MHD jet models of supernovae and of
GRBs.   Rotation remains a key parameter in stellar evolution
studies and, as for collapse dynamics, it is unlikely
that rotation will be present in the absence of magnetic fields.
Both rotation and magnetic fields must be included
self-consistently from the proto-stellar phase to collapse before
we really understand this issue.  It may not be appropriate,
for instance, to evolve a model star a considerable amount and
then add the effects of magnetic viscosity once some amount of
shear has developed.

\section{Gamma-Ray Bursts}

We ended the conference with a stimulating session on GRBs and
their possible connection to supernovae.  Tom Matheson brought
the exciting observations  by his team of supernova SN~2003dh
associated with GRB~030329.  This supernova apparently closely
resembled SN~1998bw despite the fact that GRB~030329 was a
``normal," if exceptionally close, GRB and that GRB~980425, if
associated with SN~1998bw, was rather odd.  Given the liklihood
of asymmetries and large variations in energy, it is not at all
clear why these two supernovae should be so similar.

Don Lamb summarized work on the X-ray rich bursts discovered by
HETE, concluding that the gamma-ray component might be more
collimated than previously thought and hence that the rate of
explosion of GRBs might rival that of SN Ic.  Edo
Berger, on the other hand, presented radio data on 
SN Ic that
apparently showed that rather few Type Ic could be associated
with GRBs.  Alin Panaitescu outlined his work with Pawan Kumar
that was more consistent with the ``standard" value of
collimation, but consistent for some bursts with magnetic fields
that were quite large in the early, reverse
shock phase, and that then decayed with time.  Brad Schaefer
illustrated how use of the variability/luminosity and spectral
lag/luminosity relations are both self-consistent and potentially
useful tools to provide distance estimates to GRBs and hence
to use GRBs as independent cosmological probes.

\section{Conclusions and Charge}

With SN~2003dh, we have incontrovertible evidence
that at least some GRBs are associated with spectrally identifiable
supernova explosions.  This still leaves open a raft of
fascinating questions.  What is the machine of the explosion?
My bet is that it involves rotation and magnetic fields.
With the results of Panaitescu and Kumar and of Coburn \&
Boggs (2003), there are new suggestions that the magnetic field
may not just be required to produce synchrotron radiation, but
may be dynamically important in producing the burst.  If (long)
GRBs are routinely associated with the collapse of massive stars,
how does the burst and associated magnetic field get out of the
star?  Another important question, touched on above,  is what
fraction of Type Ic supernovae make GRBs?  If GRBs come from massive
stars, why do we not see evidence for winds and shells, dense
circumstellar media, in every burst?  Must all (again long) GRBs
be associated with the formation of black holes, or can some be
associated with neutron stars?  Do we expect the distribution of
rotation rates of stars to vary over the history of the Universe,
and if so, should there be some impact on the rate of production
of GRBs with redshift?

These are important questions, but there are fundamental issues
lying at the core of all of them.  That leads to my charge to
attendees of the meeting and to readers of these proceedings.
{\it Go thee forth and think about rotation and magnetic fields!}

\bigskip

{\it Acknowledgements:}  I again express my gratitude to 
Peter H\"oflich and Pawan Kumar for the work they did on this
meeting and to all my colleagues who attended and said embarassingly
nice things.  This work was supported in part by
NSF AST-0098644 and by NASA NAG5-10766.


\end{document}